\documentclass[aps,pre,showpacs,nofootinbib,amsmath,superscriptaddress,preprintnumbers,reprint]{revtex4-2}
\usepackage{graphicx}
\usepackage{dcolumn}
\usepackage{amsmath,amssymb}
\usepackage{bbold}
\usepackage{color}
\usepackage[dvipsnames]{xcolor}
\definecolor{darkblue}{rgb}{0,0,0.6}
\usepackage[colorlinks,linkcolor=darkblue,citecolor=darkblue,urlcolor=darkblue]{hyperref}
\usepackage{soul}
\definecolor{mygreen}{rgb}{0.0,0.55,0.3}

\renewcommand{\epsilon}{\varepsilon}

\newcommand{\ee}{\text{e}}

\newcommand{\br}{\text{\bf r}}
\newcommand{\be}{\text{\bf e}}

\newcommand{\bk}{\text{\bf k}}

\newcommand{\bv}{\text{\bf v}}

\newcommand{\mC}{\mathcal{C}}
\newcommand{\mR}{\mathcal{R}}
\newcommand{\mL}{\mathcal{L}}
\newcommand{\fg}[1]{\textcolor{black}{{#1}}}

\begin{document}

\title{Monte Carlo simulations of glass-forming liquids beyond Metropolis} 

\author{Ludovic Berthier}

\affiliation{Laboratoire Charles Coulomb (L2C), Université de Montpellier \& CNRS (UMR 5221), 34095 Montpellier, France}

\affiliation{Gulliver, UMR CNRS 7083, ESPCI Paris, PSL Research University, 75005 Paris, France}

\author{Federico Ghimenti}

\affiliation{Laboratoire Mati\`ere et Syst\`emes Complexes (MSC), Université Paris Cité  \& CNRS (UMR 7057), 75013 Paris, France}

\author{Fr\'ed\'eric van Wijland}

\affiliation{Laboratoire Mati\`ere et Syst\`emes Complexes (MSC), Université Paris Cité  \& CNRS (UMR 7057), 75013 Paris, France}

\date{\today}

\begin{abstract}
Monte Carlo simulations are widely employed to measure the physical properties of glass-forming liquids in thermal equilibrium. Combined with local Monte Carlo moves, the Metropolis algorithm can also be used to simulate the relaxation dynamics, thus offering an efficient alternative to molecular dynamics. Monte Carlo simulations are however more versatile, because carefully designed Monte Carlo algorithms can more efficiently sample the rugged free energy landscape characteristic of glassy systems. After a brief overview of Monte Carlo studies of glass-formers, we define and implement a series of Monte Carlo algorithms in a three-dimensional model of polydisperse hard spheres. We show that the standard local Metropolis algorithm is the slowest, and that implementing collective moves or breaking detailed balance enhances the efficiency of the Monte Carlo sampling. We use time correlation functions to provide a microscopic interpretation of these observations. Seventy years after its invention, the Monte Carlo method remains the most efficient and versatile tool to compute low-temperatures properties in supercooled liquids. 
\end{abstract}

\maketitle

\section{Introduction: Monte Carlo simulations of glass-formers}

Understanding how glasses are formed upon slowly supercooling liquids and predicting their physical properties remain broad and important research topics~\cite{ediger1996supercooled,cavagna2009supercooled,berthier2016facets}. From a fundamental viewpoint, one would like to obtain a description of these emerging properties starting from the sole knowledge of how microscopic degrees of freedom move and interact~\cite{gotze2009complex, berthier2016facets,cavagna2009supercooled, berthier2011theoretical, parisi2020theory}. This statistical mechanics program is still incomplete, because fluctuations are typically hard to control analytically in disordered systems in finite dimensions~\cite{parisi2020theory}. Computer simulations thus represent a central tool, since finite dimensional models can be simulated with an exquisite control of the microscopic interactions leading to glass formation~\cite{berthier2023modern}.  

Although Monte Carlo~\cite{metropolis1953equation} and Molecular Dynamics~\cite{alder1959studies} were invented just a few years apart, computer studies of the glass transition in supercooled liquids have been largely dominated by Molecular Dynamics~\cite{barrat1991molecular}. This can be justified by the fact that the glass transition is primarily a kinetic phenomenon where the dramatic slowing down of the dynamics upon cooling eventually leads to the emerging solidity of the amorphous solid~\cite{ediger1996supercooled}. Instead, the Monte Carlo approach was devised to provide accurate numerical estimates of configurational averages that involve an equilibrium sampling of the configuration space~\cite{metropolis1953equation,newman1999monte,landau2021guide,frenkel2001understanding}. Thus, Monte Carlo simulations of particle systems were used to study the static structure and thermodynamic properties of liquids, thereby usefully accompanying the development of liquid state theories~\cite{hansen2013theory}. In supercooled liquids, however, the structure evolves very little as the system approaches the glass transition, at least at the level of two-point density correlation functions~\cite{gotze2009complex,berthier2011theoretical,royall2015role,tarjus2005frustration}.  

The dominance of molecular dynamics studies receded over the last two decades for at least two reasons. First, it is now understood that there is more to the structure and thermodynamics of supercooled liquids than two-body functions~\cite{parisi2020theory,royall2015role}. More complex correlation functions have been shown to be relevant~\cite{bouchaud2004on,biroli2008thermodynamic,berthier2012static}, while configurational entropy~\cite{sciortino1999inherent,sastry2001relationship,berthier2019configurational} and constrained free energies~\cite{Franz1997phase,berthier2014novel,guiselin2022statistical} are also useful indicators of the glass transition. These call for the development of efficient Monte Carlo techniques to probe configurational averages in equilibrium conditions. Second, it has been realized that details of the microscopic dynamics do not affect the glassy dynamics at long times~\cite{gleim1998how}. In particular, local Monte Carlo moves lead to dynamic relaxation pathways that are equivalent to the ones explored by the Molecular Dynamics~\cite{berthier2007monte,berthier2007revisiting,berthier2007efficient}. Thus, local Monte Carlo dynamics has become an efficient alternative to Molecular Dynamics to sample not only static but also dynamic properties of supercooled liquids in the canonical ensemble~\cite{berthier2023modern}. 

A direct consequence of the dynamic equivalence between local Monte Carlo and Molecular Dynamics is that sampling the equilibrium distribution becomes very difficult in both approaches when temperature decreases because structural relaxation times, which are good indicators of sampling efficiency and ergodic exploration, grow rapidly as the glass transition is approached. For the reasons given above, algorithms that can speedup equilibration and equilibrium sampling in deeply supercooled liquids are very much needed and sought after. Within the Monte Carlo approach, the price to pay is that the dynamic information is often lost as algorithms take unphysical shortcuts in  configuration space to achieve a faster sampling~\cite{newman1999monte,landau2021guide,frenkel2001understanding}. This is maybe not as bad as it sounds, since samples that are efficiently equilibrated at very low temperatures can still be used as initial conditions for Molecular Dynamics or local Monte Carlo simulations~\cite{scalliet2022thirty}.  

Walking in the footsteps of the Metropolis algorithm, several directions can already be followed to improve the sampling efficiency of Monte Carlo simulations of supercooled liquids~\cite{barrat2023computer}. One could invent more complicated particle moves, going beyond the small single particle translational move. A cluster algorithm was designed for dense systems of hard disks that proposes moves involving a large number of particles~\cite{dress1995cluster}. Another example is the swap Monte Carlo algorithm which exchanges the positions of randomly chosen pairs of particles. \fg{Swap Monte Carlo was first proposed in the context of simulations of multi-component solids to study their melting transition~\cite{kranendonk1991computer}, and was only much later introduced as a method for supercooled liquids and glasses~\cite{grigera2001fast}. Its efficiency has now been improved and demonstrated across a broad range of systems and spatial dimensions~\cite{ninarello2017models,berthier2019efficient,berthier2019bypassing,parmar2020ultrastable,alvarez2023simulated}.} Another generic strategy within the Metropolis realm is to add additional degrees of freedom to the system, such as temperature, to explore a larger phase space. Parallel tempering is the most studied example of this family~\cite{hukushima1996exchange}, where a range of temperatures are simultaneously simulated using replicas of the system, with infrequent temperature swaps proposed between replicas~\cite{yamamoto2000replica,jung2024normalizing}.  

The versatility of Monte Carlo simulations becomes obvious when methods reaching beyond the realm of the Metropolis acceptance rule are defined. As noted in many textbooks, the Metropolis strategy that ensures detailed balance is sufficient to reach thermal equilibrium but is not necessary, as global balance (with proof of ergodicity) is sufficient~\cite{krauth2006statistical}. In practice, defining Monte Carlo algorithms that break detailed balance and ensure Boltzmann sampling is not easy, but several such algorithms have been proposed recently in different contexts~\cite{bernard2009event, turitsyn2011irreversible}. In particular, lifting Monte Carlo algorithms introduce additional degrees of freedom to ensure global balance while breaking detailed balance~\cite{chen1999lifting, diaconis2000analysis, sakai2013dynamics, sakai2016eigenvalue, vucelja2016lifting}. For dense particle systems, the event-chain Monte Carlo algorithm was first proposed for hard disks~\cite{bernard2009event, bernard2011two}, and later generalized to soft potentials~\cite{michel2014generalized}. We have recently introduced a nonequilibrium algorithm~\cite{ghimenti2024irreversible} where lifting is applied to both translational and diameter degrees of freedom, thus generalizing both the event-chain and swap Monte Carlo algorithms. 

This brief overview of Monte Carlo approaches applied to supercooled liquids suggests that more efficient algorithms are still crucially needed. Here, we build on our previous work on two-dimensional hard disks~\cite{ghimenti2024irreversible} and assess whether the proposed nonequilibrium algorithms remain efficient in three dimensions. To this end, we define and implement a similar family of Monte Carlo algorithms for a polydisperse model of hard spheres known to display glassy dynamics~\cite{berthier2016equilibrium}. We then compare their performances to reach and sample equilibrium as well as for nonequilibrium compressions towards the jamming transition. Overall, our results show that the performances of nonequilibrium swap algorithms remain very good in three dimensions.     

The manuscript is organized as follows. 
In Sec.~\ref{sec:model} we define the studied model of hard spheres. 
In Sec.~\ref{sec:algorithms}, we introduce the various Monte Carlo algorithms we analyze. 
In Sec.~\ref{sec:EOS} we confirm that all algorithms properly sample the same equilibrium Boltzmann distribution, by following the equation of state.
In Sec.~\ref{sec:tau_alpha} we compare the efficiency of all Monte Carlo algorithms.
In Sec.~\ref{sec:insights} we study various time correlation functions to provide microscopic insights into the performances of the algorithms. 
Finally, in Sec.~\ref{sec:jamming} we use these Monte Carlo algorithms to prepare very dense jammed packings of spheres. We conclude the paper in Sec.~\ref{sec:conclusion}. 

\section{Polydisperse hard sphere model}

\label{sec:model}

We simulate a polydisperse mixture of $N=1000$ hard spheres in three dimensions. The pair interaction is $V(r)=0$ when the distance between particle pairs $(i,j)$ satisfies $r_{ij}> \sigma_{ij}$ with $\sigma_{ij} = (\sigma_i + \sigma_j)/2$, and $V(r)=\infty$ otherwise. 

The diameters $\{ \sigma_i, i=1 \cdots N \}$ of the particles are drawn from a power law distribution~\cite{berthier2016equilibrium,berthier2017configurational}, $\pi(\sigma) \propto \sigma^{-3}$, with boundaries $\sigma_{\rm min}$ and $\sigma_{\rm max}$ chosen so that the polydispersity $\Delta \equiv \frac{\sqrt{\overline{\sigma^2} - \overline{\sigma}^2}}{\overline{\sigma}}$ is $\Delta\approx 23\%$, with the overline \fg{$\overline{\cdots}$} denoting an average over $\pi(\sigma)$. We study the dynamics of the system as a function of the packing fraction $\phi$ defined as 
\begin{equation}
    \phi = \frac{\pi N \overline{\sigma^3}}{6V},
\end{equation}
where $V$ is the volume of the system. Periodic boundary conditions are used. 
In Secs.~\ref{sec:tau_alpha} and \ref{sec:insights} we use the $NVT$ ensemble, where the volume $V$ is fixed. We move to the $NPT$ ensemble to perform compressions to measure the equation of state in Sec.~\ref{sec:EOS} and to arrive to the jamming point in Sec.~\ref{sec:jamming}. In that case, the volume of the simulation box is allowed to vary to maintain the pressure $P$ constant. 

We use reduced units to present our results: the packing fraction is non-dimensional, as well as the compressibility factor $Z = P / (\rho k_B T)$, with $\rho = N/V$ the number density and $k_B$ the Boltzmann constant. The average particle diameter $\overline{\sigma}$ sets the unit length. For each algorithm presented in Sec.~\ref{sec:algorithms}, we define an elementary timescale $t_{\rm move}$ such that $\tau_0 = N t_{\rm move}$ typically represents a Monte Carlo sweep   across the system. Therefore we use $\tau_0$ as the time unit when presenting time-dependent quantities and relaxation timescales.   

\fg{In Ref.~\cite{ghimenti2024irreversible}, we showed that the efficiency of all algorithms discussed in this work is insensitive to system size, so we study a single value of $N$ in the present manuscript.}

\section{Monte Carlo algorithms: Metropolis and beyond}

\label{sec:algorithms}

\begin{figure}
    \includegraphics[width=\columnwidth]{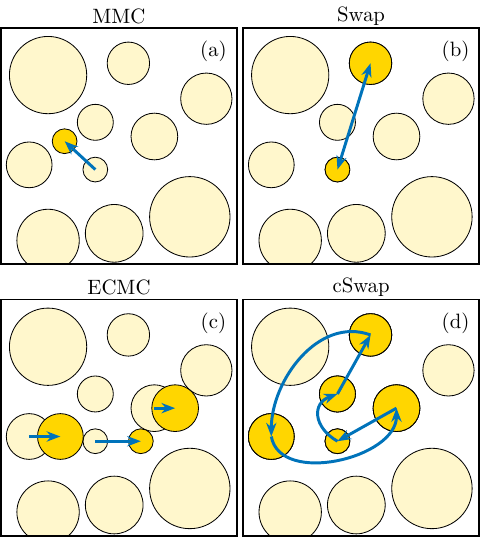}
    \caption{Elementary Monte Carlo moves. 
    (a) In the Metropolis Monte Carlo (MMC) algorithm, randomly chosen particles perform small random displacements, which are accepted according to the Metropolis rule. (b) In the Swap algorithm, randomly chosen pairs of particles exchange their positions with an acceptance probability given by the Metropolis rule. (c) In the Event Chain Monte Carlo (ECMC) , directed chains of particle displacements are implemented. This violates detailed balance, but preserves the Boltzmann distribution. (d) In the collective Swap (cSwap) algorithm, collective swap moves are implemented. This violates detailed balance but preserves the Boltzmann distribution.}
    \label{fig:cartoon} 
\end{figure}

In this section we introduce the Monte Carlo algorithms implemented to simulate the hard sphere model presented in Sec.~\ref{sec:model}. Following~\cite{ghimenti2024irreversible}, we carefully introduce a unit of time $t_\text{move}$ for each algorithm, from which a quantitative comparison of their performances can be achieved. In Fig.~\ref{fig:cartoon} we sketch the various types of Monte Carlo moves described in this section. 

\subsection{Metropolis algorithms: Local moves (MMC) and Swap algorithm}

\label{sub:metropolis}

In the simplest version of the Metropolis Monte Carlo~\cite{metropolis1953equation} (MMC) algorithm, during an elementary time $t_\text{move}$ a particle is randomly and uniformly selected, and a small random displacement is proposed, see Fig.~\ref{fig:cartoon}(a). The displacement vector is uniformly drawn from a cube of linear size $\delta$ centered around the particle. The proposed displacement is accepted if it does not generate any overlap with the neighboring particles, which is the consequence of using the Metropolis rule for the hard core potential $V(r)$ defined in Sec.~\ref{sec:model}. Therefore, the MMC algorithm is local and satisfies detailed balance, thus leading to an equilibrium sampling of the Boltzmann distribution. For the simulations presented in this work we choose $\delta = 0.115\overline{\sigma}$, obtained by optimizing the relaxation time of the system at intermediate packing fraction~\cite{berthier2007monte}.

In the swap Monte Carlo algorithm~\cite{grigera2001fast, ninarello2017models}, which we simply call `Swap', pairs of particles can perform swap moves. A single swap move takes place during a time interval $t_\text{move}$. During a swap move, a pair of particles is randomly and uniformly selected, and their positions are exchanged if this generates no overlap among the spheres, see Fig.~\ref{fig:cartoon}(b). This acceptance rule is again the consequence of the Metropolis rule for the hard core potential $V(r)$ in the case of a proposed particle swap. The sketch in Fig.~\ref{fig:cartoon}(b) shows that it is equivalent to considering that the two swapped particles exchange their diameters at fixed position, or exchange their positions at fixed diameters. While the latter view is more natural, we adopt the former because it allows a simpler definition of single-particle time correlation functions~\cite{ninarello2017models}, as studied in Sec.~\ref{sec:insights}. In the Swap algorithm, we alternate randomly between sets of $N$ swap moves and sets of $N$ MMC moves. A set of $N$ swap moves is performed with probability $p_\text{Swap}$, while a set of $N$ MMC moves is performed with probability $1-p_\text{Swap}$. Following previous work~\cite{berthier2016equilibrium, ninarello2017models}, we take $p_\text{Swap}=0.2$, which optimizes the relaxation time for the decay of the self-intermediate scattering function, see Eq.~\eqref{eq:Fs}, and definitions below.

\subsection{Beyond Metropolis: Event-chain (ECMC) and collective swap (cSwap)}

Both the local Metropolis and the Swap algorithms in Sec.~\ref{sub:metropolis} obey detailed balance, which is a sufficient condition to achieve a proper sampling of the Boltzmann distribution in the steady state. Both for translational moves and diameter swaps, the design of algorithms that break detailed balance while preserving the Boltzmann distribution in their stationary state is possible. Such strategy goes beyond the seminal Metropolis approach~\cite{metropolis1953equation}.  We now present two possible dynamics for the hard sphere model. 

In the original ECMC algorithm~\cite{bernard2009event}, the configuration space of the particles' positions $\br^N$ is lifted to an extended space $\{ \br^N,i,\bv \}$, which now includes the label $i$ of an active particle and a direction of motion $\bv$. During a time $t_\text{move}$, the active particle $i$ moves along the direction $\bv$, until it collides with a neighbor $j$. When this happens, a lifting event takes place, and the activity label changes from $i$ to $j$. In this way, after a time $n t_\text{move}$, a directed chain of $n$ particles has been displaced along the direction $\bv$, see Fig.~\ref{fig:cartoon}(c). When the individual displacements of the particles involved in a chain add up to a total length $\ell$, the activity label is uniformly resampled among the $N$ particles, while the self-propulsion direction $\bv$ is uniformly resampled in the set $\{\be_x, \be_y, \be_z\}$ in order to restart another chain. Contrary to Metropolis algorithms, there is no need to satisfy any acceptance rule, which renders the algorithm rejection-free. Also, while detailed balance is broken (chain moves cannot be reverted), the Boltzmann distribution is the solution of the global balance condition~\cite{bernard2009event}. The efficiency of ECMC is roughly independent on the chain length, as long as $\ell$ is of the order of a few particle diameters and smaller than the linear size of the system. \fg{Since this choice is not critical, we choose $\ell = 2\overline{\sigma} \sim 0.2 L$, and do not vary $\ell$ further~\cite{ghimenti2024irreversible}.}

It is of course possible to combine ECMC with Swap to define the SwapECMC algorithm~\cite{ghimenti2024irreversible}. To this end,  we choose between an ECMC move which builds a chain of length $\ell$ and $N$ swap moves with a probability $p_\text{SwapECMC}$ and $(1-p_\text{SwapECMC})$, respectively. We take $p_\text{SwapECMC}=0.2$. 

In the cSwap algorithm, the lifting framework used in ECMC is implemented in diameter space, rather than in position space~\cite{ghimenti2024irreversible}. A one dimensional array storing the label of the particles in order of increasing diameters is created at the beginning of a simulation. We define the operators $\mL,\mR$, that, for a given particle $i$, select respectively the particle to the left or to the right of $i$ along the ordered array, applying periodic boundary conditions. A configuration $\mC \equiv \{ \br^N, \sigma^N, i\}$ of the system includes all positions $\br^N$, diameters $\sigma^N$, and an activity label $i$, which denotes an active particle. During  a time $t_\text{move}$, the largest possible diameter $\sigma_j$ (corresponding to particle $j$) that can be reached by particle $i$ without generating overlaps is identified and particle $i$ is inflated: $\sigma_i \leftarrow \sigma_j$. The $n_s$ non-active particles with a diameter comprised between $\sigma_i$ and $\sigma_j$ (including $j$), perform a small deflation, moving one step to the left in the ordered array of particle diameters: $\sigma_{\mL j} \leftarrow \sigma_{\mL^2 j}, \ldots,\sigma_{\mL^{n_s-1} j} \leftarrow \sigma_{\mL^{ns} j}$. Finally, the activity label switches from $i$ to $\mL i$. To warrant ergodicity, the activity label is resampled uniformly among all the particles with probability $1 - \frac{1}{N_\text{cSwap}}$. Overall, this results in a collective swap move, as sketched in Fig.~\ref{fig:cartoon}(d). The algorithm is rejection-free, it does not invoke the Metropolis acceptance rule, and it breaks detailed balance. In a simulation, a set of $N_\text{cSwap}$ moves is performed with probability $p_\text{cSwap}$, while with probability $1 - p_\text{cSwap}$ we perform a set of $N$ local translational Metropolis moves. We fix the probability $p_\text{cSwap}=0.2$, and we optimize the number of Swap moves $N_\text{cSwap}$ at a density $\phi=0.624$, where equilibrated samples are obtained with a modest effort using the cSwap algorithm. We thus set $N_\text{cSwap}=500$.

We recently showed~\cite{ghimenti2024irreversible} that for the hard sphere interaction $V(r)$, the stationary distribution of the cSwap algorithm, once marginalized over the lifted degrees of freedom, is the Boltzmann distribution. By construction, the particle size distribution is preserved by an elementary cSwap move. {In addition, the collective swap moves are performed with probability $p_\text{cSwap}$, while elementary translational moves as in MMC are performed with probability $(1-p_\text{cSwap})$}.  

Finally, ECMC and cSwap can be combined in the cSwapECMC algorithm~\cite{ghimenti2024irreversible}. In this algorithm, we perform $N_\text{cSwap}$ cSwap moves with probability $p_\text{cSwapECMC}$, and we otherwise displace a chain of length $\ell$ using the ECMC procedure with probability $1 - p_\text{cSwapECMC}$. We set $p_\text{cSwapECMC}=0.2$.

\section{Equilibrium sampling: Equation of state}

\label{sec:EOS}

All the algorithms proposed in Sec.~\ref{sec:algorithms} can be shown to lead to a proper sampling of the same Boltzmann distribution in the steady state. Thus, equilibrium sampling is preserved even if the algorithms that break detailed balance are, technically, evolving out of equilibrium. To numerically demonstrate that this is the case, we perform slow compressions of the hard sphere system using MMC, ECMC, Swap, and cSwap starting from a relatively dilute fluid state at $\phi=0.53$. To slowly compress the system we introduce Monte Carlo moves of the volume~\cite{frenkel2001understanding}. 

In a volume move, a volume change $\delta V$ is sampled uniformly in an interval of width $\delta V_\text{max}$ centered around $0$. The new configuration proposed for the system is then a box of volume $V+\delta V$, with particle positions rescaled by a factor $s = \left(\frac{V+\delta V}{V}\right)^{1/3}$. This configuration is accepted with a Metropolis acceptance rate, tuned to ensure sampling of the $NPT$ ensemble. The new configuration is accepted with probability $\ee^{-\beta P\delta V + N\log s}$ if no overlap among the spheres are generated by the volume change. 

\begin{figure} 
    \includegraphics[width=\columnwidth]{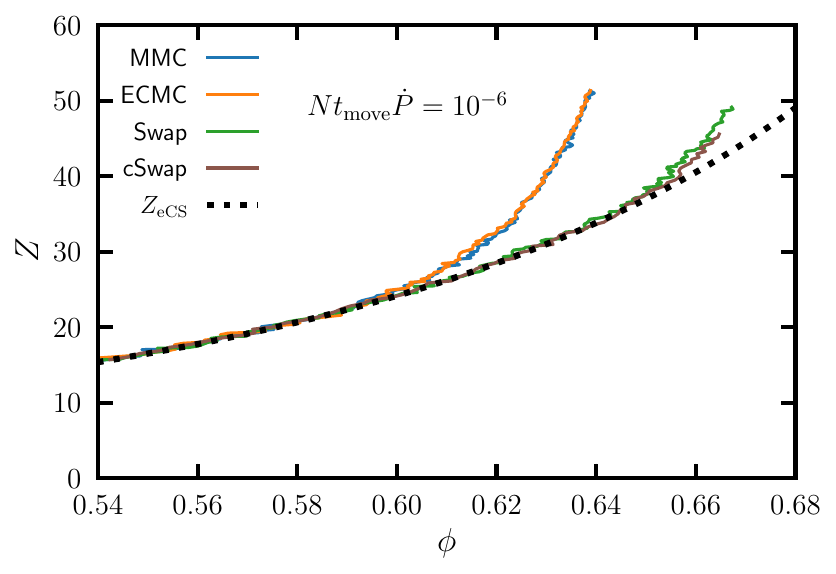}
    \caption{Equation of state $Z=Z(\phi)$ obtained from four different algorithms during slow compressions. The black dashed line is the Carnahan-Starling formula $Z_{\rm CS}$ from Eq.~(\ref{eq:EOS}).}
    \label{fig:eos}
\end{figure}

Volume moves are introduced in a different way for each algorithm. For MMC, Swap and cSwap, we replace with probability $1-\frac{1}{N}$ a single translational move with a volume move. This choice is justified by the necessity to prevent the algorithm from becoming periodic~\cite{frenkel2001understanding}. In ECMC, the presence of a finite chain length ensures that the algorithm is unbiased~\cite{michel2014generalized}. The volume moves are thus implemented by replacing the displacement of a chain of length $\ell$ with a volume move with probability $1/2$.

We use a slow compression rate, $\dot P = 10^{-6}$, for all algorithms. This is small enough that the system can easily reach the Boltzmann distribution in the fluid state. However, when the packing fraction becomes large, glassy dynamics emerges and for each algorithm there will exist a packing fraction above which thermal equilibrium can no longer be achieved.

In Fig.~\ref{fig:eos} we present the evolution of the compressibility factor $Z$ with the packing fraction $\phi$ for four algorithms. As expected, \fg{all the algorithms} produce the same equation of state $Z=Z(\phi)$ in the stationary fluid state. The numerical results are in good agreement with the extension to polydisperse systems of the empirical Carnahan-Starling formula~\cite{boublik1970hard}:
\begin{equation}
    Z_\text{CS} = \frac{1}{1-\phi} + \frac{3\phi}{(1-\phi)^2}\frac{\overline{\sigma}\overline{\sigma^2}}{\overline{\sigma^3}} + \frac{\phi^2(3-\phi)}{(1-\phi)^3}\frac{\overline{\sigma^2}^3}{\overline{\sigma^3}^2},
    \label{eq:EOS}
\end{equation}
which requires moments of the particle size distribution as input. The mutual agreement between algorithms confirms that they all probe the same stationary distribution. 

In addition, for each algorithm we observe a packing fraction above which clear deviations from the stationary equation of state are observed. This happens near $\phi \approx 0.60$ for MMC and ECMC, and near $\phi \approx 0.64$ for Swap and cSwap. At these packing fractions, the system fails to relax within the compression timescale and the measured equation of state differs from the equilibrium result. Consistently with previous results~\cite{berthier2016equilibrium}, the introduction of diameter moves considerably extends the stationary regime. There are subtle physical differences between MMC and ECMC and \fg{between Swap and cSwap} that go beyond the mere reduced convergence time, but further explorations are required to reveal them, as we do in the next Section.

\section{Relative efficiency of Monte Carlo algorithms}

\label{sec:tau_alpha}

\begin{figure}[t]
    \includegraphics[width=\columnwidth]{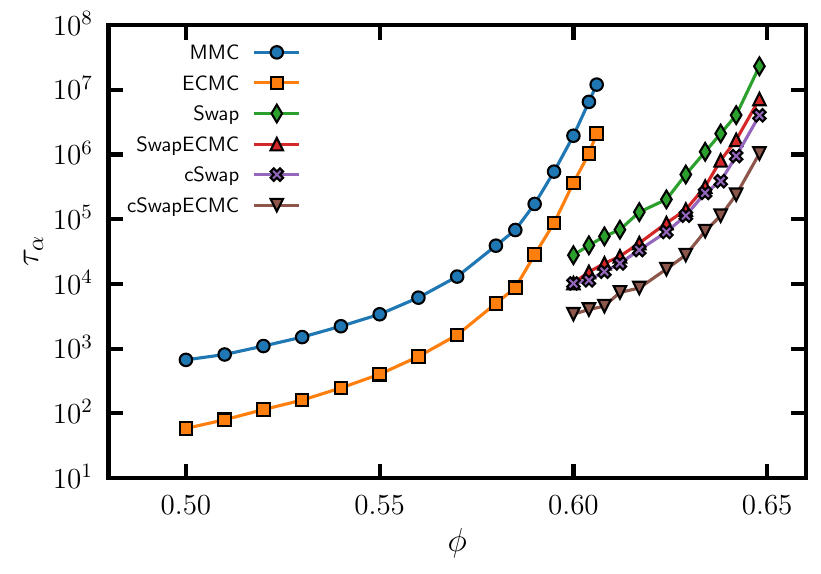}
    \caption{Evolution of the relaxation time $\tau_\alpha$ for MMC, ECMC, Swap, SwapECMC, cSwap, cSwapECMC algorithms as a function of the packing fraction $\phi$.}
    \label{fig:tau_alpha}
\end{figure}

To study the relative efficiency of the various Monte Carlo algorithms, we need to define a relevant timescale that captures the rate at which the configuration space is explored. To this end, \fg{we define the self-intermediate scattering function as}
\begin{equation}\label{eq:Fs}
    F_s(\bk,t) \equiv \frac{1}{N}\sum_i\left\langle \ee^{-i\bk\cdot\Delta\br_i(t)}\right\rangle,
\end{equation}
with $\Delta\br_i(t) \equiv \br_i(t) - \br_i(0) - \Delta\br_\text{com}(t)$, and $\Delta\br_\text{com}$ is the displacement of the center of mass of the system over the time interval $[0,t]$, and the brackets $\langle \ldots \rangle$  indicate an ensemble average over the initial conditions and the dynamics. The modulus of the wavevector $\bk$ is chosen to correspond to the first peak of the static structure factor, $\lvert \bk \rvert \approx \frac{2\pi}{\overline{\sigma}}$. From the time decay of $F_s$, the relaxation time $\tau_\alpha$ is defined as $F_s(\bk,t=\tau_\alpha) \equiv \ee^{-1}$. 
\begin{figure}[t]
    \includegraphics[width=\columnwidth]{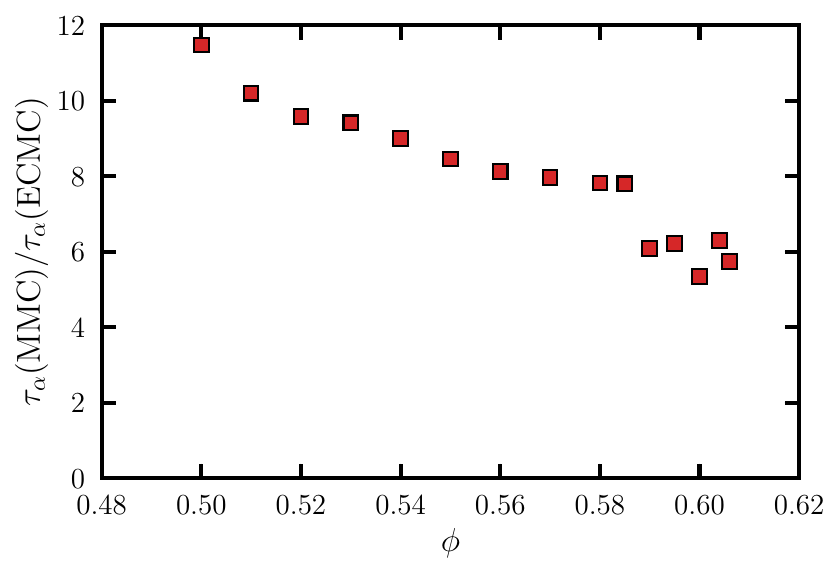}
    \caption{Relative efficiency of nonequilibrium ECMC versus equilibrium MMC. The speedup due to collective chain moves decreases slowly with the density.
    \label{fig:MMC_vs_ECMC}} 
\end{figure}

For a given packing fraction, we prepare equilibrated samples by performing simulations in the $NVT$ ensemble using the cSwapECMC algorithm for a duration of $\sim 10 \tau_\alpha$. Using these equilibrated samples as initial conditions, we then perform $NVT$ simulations using the six different algorithms (MMC, ECMC, Swap, SwapECMC, cSwap, cSwapECMC) in the stationary regime. We average $F_s(\bk,t)$ over 40 independent runs for ECMC and MMC, and over 20 independent runs for the other algorithms. We also average over all the wavevector of the same modulus. From these curves, we extract $\tau_\alpha$ by means of an interpolation on a logarithmic time axis. It is known that $\tau_\alpha$ is a good indicator of how fast structural relaxation occurs at large $\phi$. Therefore the relative efficiency of the Monte Carlo algorithms can be obtained by comparing how fast the various algorithms decorrelate the structure of the system. 

The relaxation times for the different algorithms as a function of the packing fraction $\phi$ are shown in Fig.~\ref{fig:tau_alpha}. Clearly, the curves split into two families: one in which only translational moves are performed (MMC, ECMC) and one in which diameter moves are combined with particle translations (Swap, SwapECMC, cSwap, cSwapECMC). The latter is much more efficient that the former.

Full relaxation of the system using translational moves only becomes hard when $\phi \approx 0.61$. Although both are close, ECMC maintains an edge over MMC across the whole range of packing fractions investigated in our work. However, the gap between the two algorithms becomes smaller as we probe higher densities. This is illustrated in Fig~\ref{fig:MMC_vs_ECMC}, where the ratio of the relaxation times for the two algorithms is plotted as a function of $\phi$, reaching a factor about 6 near $\phi=0.61$.  

\begin{figure}
    \includegraphics[width=\columnwidth]{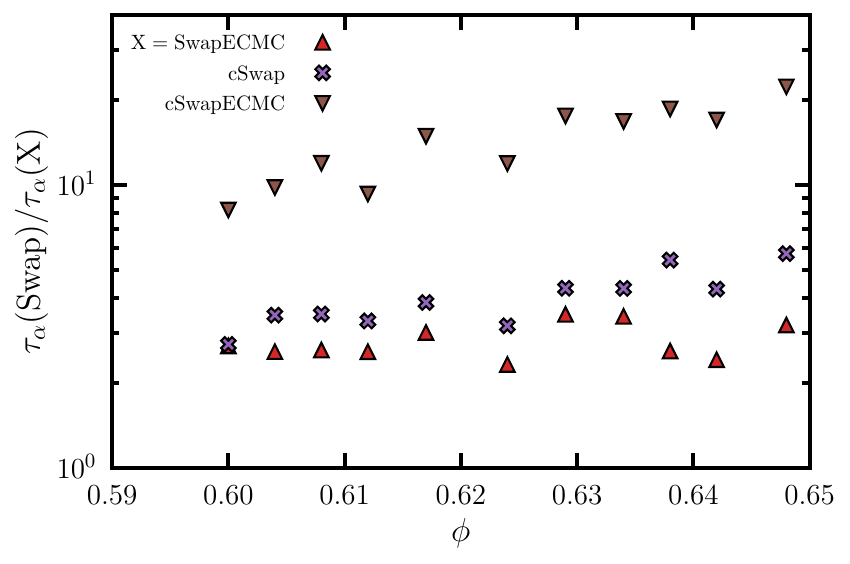}
    \caption{Efficiency of SwapECMC, cSwap and cSwapECMC relative to the Metropolis Swap algorithm, as a function of the packing fraction $\phi$. Moving beyond Metropolis systematically improves the performances of Monte Carlo algorithms at large $\phi$.}
    \label{fig:eff_swap} 
\end{figure}

When diameter swaps moves are introduced, it is possible to equilibrate samples up to much higher densities near $\phi \approx 0.65$. The relative efficiency of the different swap Monte Carlo algorithms is presented in Fig.~\ref{fig:eff_swap}. Comparing Swap to SwapECMC shows that replacing equilibrium local particle translations with collective chain moves marginally improves the efficiency, reducing the relaxation time by a factor $3$, which does not depend sensibly on $\phi$. On the other hand, the efficiency of cSwap increases over that of Swap with density, reaching a factor $6$ at $\phi=0.648$. Finally, cSwapECMC combines the advantage of the nonequilibrium nature of both diameter exchanges and translational moves, reaching a speedup of $20$ at $\phi=0.648$, suggesting that the two speedups are nearly multiplicative. 

The results obtained in this Section are very similar to our recent findings for the two-dimensional hard-disk system~\cite{ghimenti2024irreversible}.

\section{Insights from microscopic dynamics}

\label{sec:insights}

To gain physical insights into the relative performances of the various Monte Carlo algorithms, we measure their respective time correlation functions and compare their behaviors.  

\subsection{Mean-squared displacement}

\label{sec:msd}

\begin{figure}
    \includegraphics[width=\columnwidth]{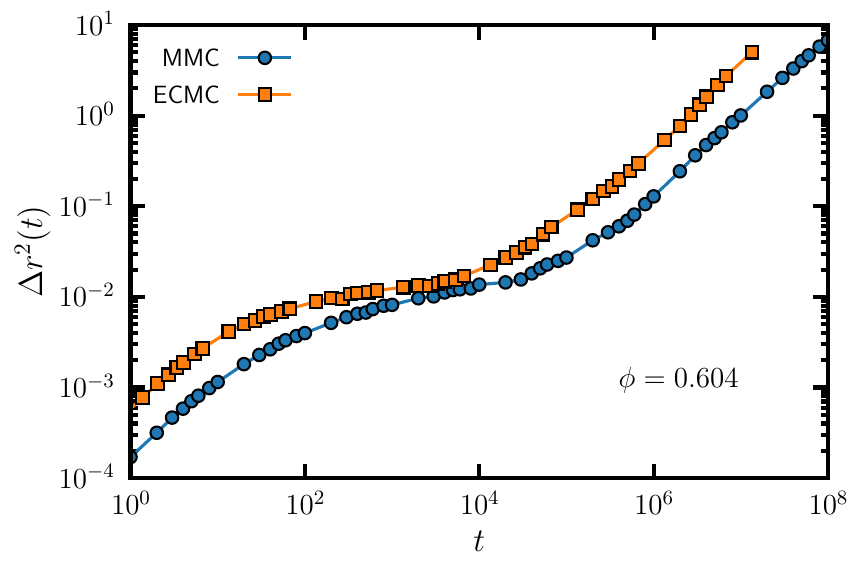}
    \caption{The mean-squared displacement, Eq.~(\ref{eq:MSD}), for MMC and ECMC at $\phi=0.604$ are very similar, apart from a global time rescaling.}
    \label{fig:msd_MMC_vs_ECMC}
\end{figure}

The mean-squared displacement $\Delta r^2(t)$ of the particles relative to the center of mass is defined as
\begin{equation}
    \Delta r^2(t) \equiv \frac{1}{N}\sum_i \langle \left[ \Delta\br_i(t)\right]^2\rangle.
\label{eq:MSD}
\end{equation}
The behavior of $\Delta r^2(t)$ for ECMC and MMC is compared at high packing fraction in Fig.~\ref{fig:msd_MMC_vs_ECMC}. Both curves show a very similar time dependence. They display a short time diffusive behavior followed by a plateau at intermediate times, reflecting particle caging. They eventually display diffusive behavior again at long times. The plateau height is similar in both curves. These curves seemingly only differ by a global rescaling of the time, implying that both the short-time and long-time diffusion constants are accelerated by a similar amount when event-chain translational moves are introduced. The similarity between these two functions suggests that with the ECMC algorithm the system follows dynamic relaxation pathways that are similar to the ones of followed when resorting to MMC, up to the fact that these paths are explored faster.  

\begin{figure}    
    \includegraphics[width=\columnwidth]{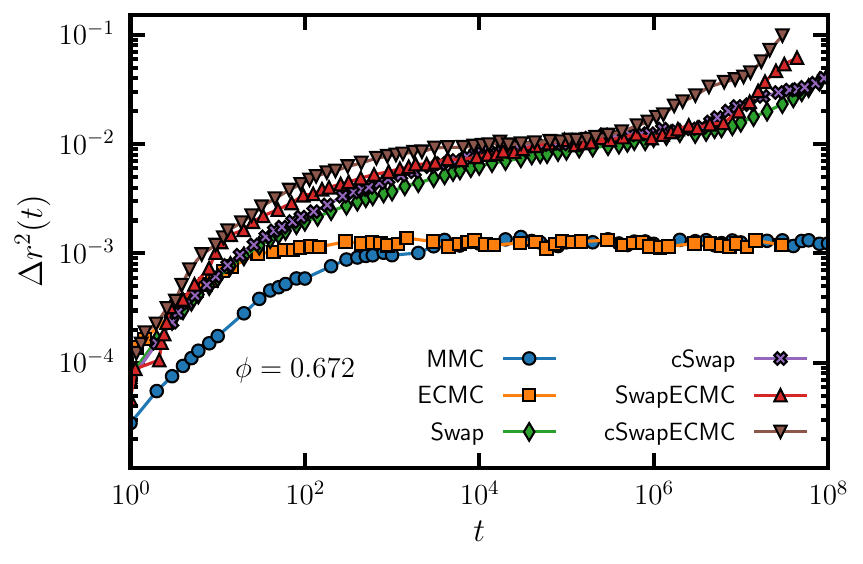}
    \caption{Mean-squared displacement, Eq.~(\ref{eq:MSD}) at high density $\phi=0.672$ for MMC, ECMC, Swap, cSwap and cSwapECMC. Swap moves increase the plateau height corresponding to caged particle motion.}
    \label{fig:msd_high_phi}
\end{figure}

We also investigate the mean-squared displacement at a much larger packing fraction, $\phi=0.672$, where equilibration  was achieved using the cSwapECMC algorithm which is the most efficient. We use 15 independent equilibrated configurations obtained via cSwapECMC as initial conditions for trajectories run with the other algorithms. The results, shown in Fig.~\ref{fig:msd_high_phi}, are therefore representative of the stationary relaxation dynamics explored by the various algorithms over the time window of the simulations~\cite{scalliet2022thirty}.  

At this very large density, MMC and ECMC are totally unable to relax the system, and particles simply explore their cages over the entire time window, thus producing a very clear plateau in the mean-squared displacements. Again, ECMC reaches the plateau a bit faster than MMC, but the plateau values are identical. 

The situation is different for Swap, which closely follows the MMC data at very short times but reaches a plateau value that is about three times as large as that of MMC and ECMC. This suggests that particle swaps dramatically enhance the amplitude of the in-cage motion performed by the particles, as noted before~\cite{ninarello2017models}. Moving from Swap to SwapECMC produces a slightly faster approach to a plateau value that is similar in both cases. Finally, moving from SwapECMC to the most efficient cSwapECMC, we observe a similar time-dependence at short times, but the plateau height is also slightly larger, with a long-time dynamics that starts to enter the diffusive regime before all other algorithms, which is consistent with the speedup observed by measuring $\tau_\alpha$ in Section~\ref{sec:tau_alpha}.  

\subsection{Non-Gaussian parameter}

\begin{figure}
    \includegraphics[width=\columnwidth]{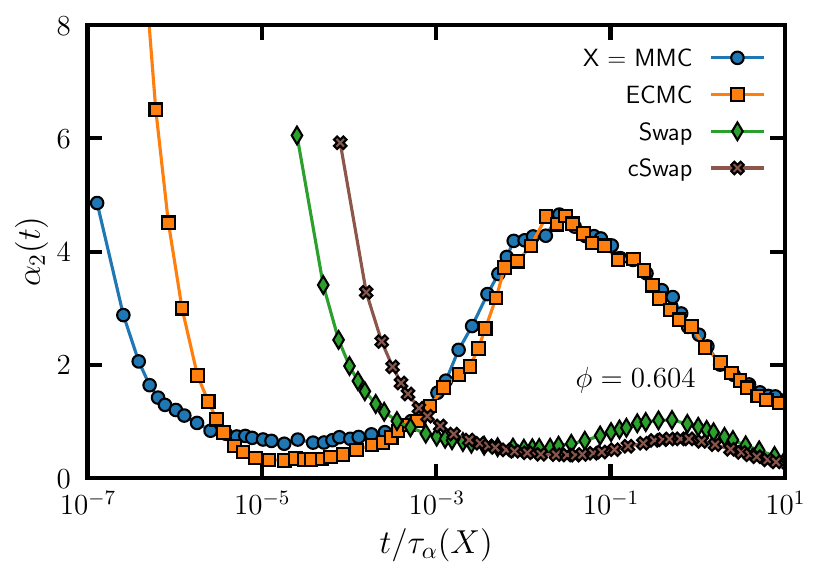}
    \caption{Non-Gaussian parameter $\alpha_2(t)$ for MMC and ECMC. for each algorithm, the time has been rescaled by the corresponding relaxation time $\tau_\alpha$.}
    \label{fig:alpha2} 
\end{figure}

The decreasing efficiency of ECMC relative to MMC at high packing fraction suggests that the collective displacement of particle chains struggle to produce displacements that are significantly different from the one produced by MMC. To confirm this picture at the level of single particle displacements, we measure the non-Gaussian parameter $\alpha_2(t)$, defined as
\begin{equation}
\alpha_2(t) \equiv \frac{3\sum_i \langle\Delta \br_i(t)^4 \rangle}{5 \sum_i \langle\Delta \br_i(t)^2\rangle^2} - 1.    
\end{equation}
This quantity measures the deviation of the probability distribution function of particle displacements from a Gaussian. It is a proxy to quantify the heterogeneity of single-particle displacements in glassy systems~\cite{rahman1964correlations,kob1997dynamical}.

A plot of $\alpha_2(t)$ for MMC, ECMC, Swap and cSwap is shown in Fig.~\ref{fig:alpha2}. In this graph, the time is rescaled by the relaxation time $\tau_\alpha$ of each algorithm. Unlike for Molecular Dynamics, the value of $\alpha_2(t)$ at very short times is non zero since both the Metropolis translations and the ECMC chain displacements are, by construction, non-Gaussians in this regime. Moreover, the value of $\alpha_2(t)$ at short times is larger for ECMC than MMC, as a reflection of the physical differences between the type of moves performed by the two algorithms. At larger times, however the two functions for ECMC and MMC, become virtually identical with a similar non-monotonic time dependence and a peak of equal amplitude near $t/\tau_\alpha \sim 0.1$. This suggests that the nature of single particle displacement at large times is similar for ECMC and MMC algorithms. This is consistent with the picture suggested by the mean-squared displacement of Sec.~\ref{sec:msd}, and the displacements fields obtained in two dimensions~\cite{ghimenti2024irreversible}. The behavior for the Swap and cSwap algorithms is instead very different. At short times, the non-Gaussian parameter has a similar value for both algorithms, since they both exploit Metropolis Monte Carlo moves for the particle translations. The non-monotonic dependence near the relaxation time is much weaker, with a much smaller maximum. Upon rescaling time by $\tau_\alpha$, the curves for Swap and cSwap are close, with cSwap being even smaller than Swap. The impressive change in the behavior of the non-Gaussian parameter upon introducing Swap moves is consistent with the analysis of the mean-squared displacement in Sec.~\ref{sec:msd}, and suggests that swap moves considerably decrease the dynamic heterogeneity~\cite{ninarello2017models}.  

\subsection{Diameter dynamics}

\begin{figure}
    \includegraphics[width=\columnwidth]{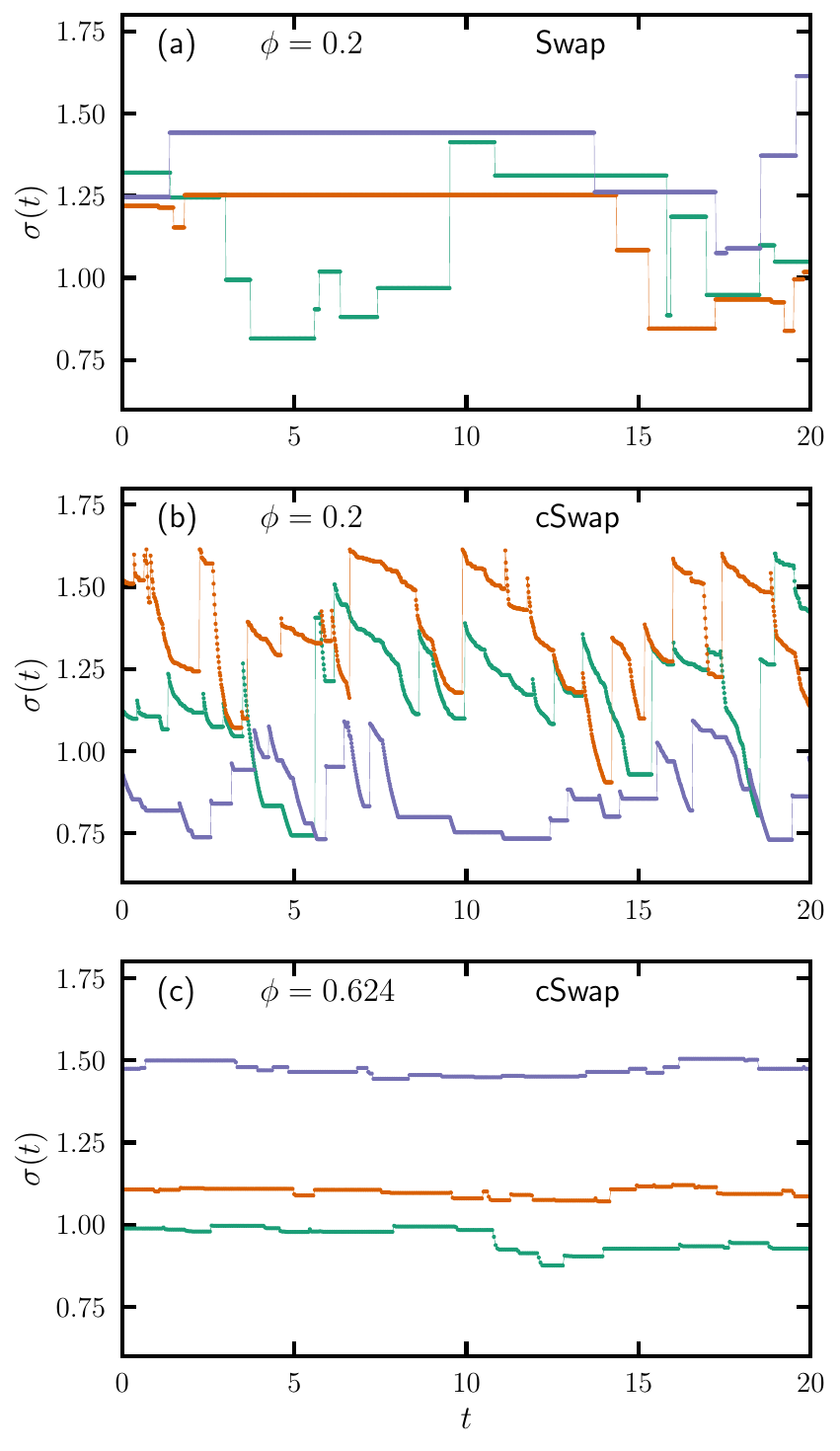}
    \caption{\fg{Time series of the diameter of three tagged particles for (a) Swap and (b) cSwap dynamics in the dilute fluid. In (c) we show the cSwap time series for a much denser system. The continuous time random walk picture for diameter dynamics in Swap becomes directed as a result of the collective swap moves in cSwap, but diameter jumps become smaller at larger density.}}
    \label{fig:sigma_t_cswap_vs_Swap_highphi}
\end{figure}

To understand the effect of collective swap moves introduced in cSwap, we focus on the dynamics of the particle diameters. As explained in Sec.~\ref{sec:algorithms}, swap Monte Carlo moves can be interpreted as providing a time dependence to the particle diameters, promoting $\sigma_i$ from a constant to a fluctuating degree of freedom~\cite{berthier2019efficient}. 

In Fig.~\ref{fig:sigma_t_cswap_vs_Swap_highphi} we show short time series of the diameter $\sigma_i(t)$ for three randomly-selected tagged particles at a relatively low packing fraction, $\phi=0.2$ for both Swap and cSwap algorithms. Since the particles diameters do not change when translational moves are performed, in the time series only swap moves are taken into account. The time $t$ in Fig.~\ref{fig:sigma_t_cswap_vs_Swap_highphi} thus counts the number of cSwap or Swap moves performed. Under Swap dynamics shown in Fig.~\ref{fig:sigma_t_cswap_vs_Swap_highphi}(a), the diameters perform a continuous time random walk. There are moments where the diameter is constant either because particle $i$ is not selected, or because swap moves are not accepted. When the diameter changes as a result of a successful swap, its value can increase or decrease, so that at long times the diameter performs a diffusive exploration of the particle size distribution $\pi(\sigma)$.   

The dynamics is very different under cSwap dynamics, where a form of directed motion is apparent, see Fig.~\ref{fig:sigma_t_cswap_vs_Swap_highphi}(b). Each time series now display a succession of continuous decrease of the diameter, interrupted by sharp inflations. These upwards jumps correspond to the situation where the tagged particle has inherited the activity tag and a collective swap move is performed from this particle. In between the sudden expansion steps, the diameter undergoes small deflations and moments of stasis. The deflations take place when the particle participate in a collective swap move, while being inactive. The moments of stasis happen when the particle is neither active nor involved in a collective swap. \fg{When density becomes larger, as shown in Fig.~\ref{fig:sigma_t_cswap_vs_Swap_highphi}(c), the moments of stasis become longer, and the amplitude of the expansion steps decrease.}

\begin{figure}
    \includegraphics[width=\columnwidth]{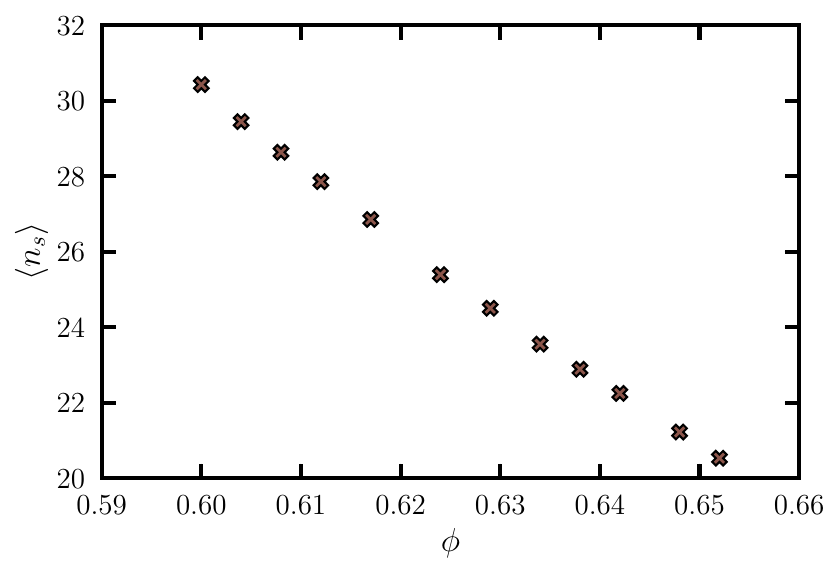}
    \caption{Evolution of $\langle n_s \rangle$, the average number of particles involved in collective swap moves with the packing fraction.}     
    \label{fig:cSwap_cascade}
\end{figure}

Unlike ECMC, within cSwap there is no need to prescribe a length for the collective moves. Instead, the number of particles $n_s$ participating to a collective move is set by the maximal possible expansion of the active particle. Clearly, this expansion must be a function of the packing fraction. To analyze this, we measure $\langle n_s\rangle$, the average number of particles that participate in collective swaps, see Fig.~\ref{fig:cSwap_cascade}. While $\langle n_s \rangle\sim N$ in the dilute limit $\phi \to 0$, this number decreases when the packing fraction increases. We observe $\langle n_s\rangle  \approx 190$ at $\phi=0.20$ and a much smaller number, $\langle n_s \rangle \approx 30$ near $\phi=0.6$ where MMC becomes unable to equilibrate. Near $\phi=0.65$ where cSwap also becomes very slow, the collective swap moves still involve a significant number of particles, $\langle n_s \rangle \approx 20$. Recalling that in addition collective swaps are never rejected, the enhanced efficiency of cSwap over swap is easily accounted for by these observations.  

\begin{figure}
    \includegraphics[width=\columnwidth]{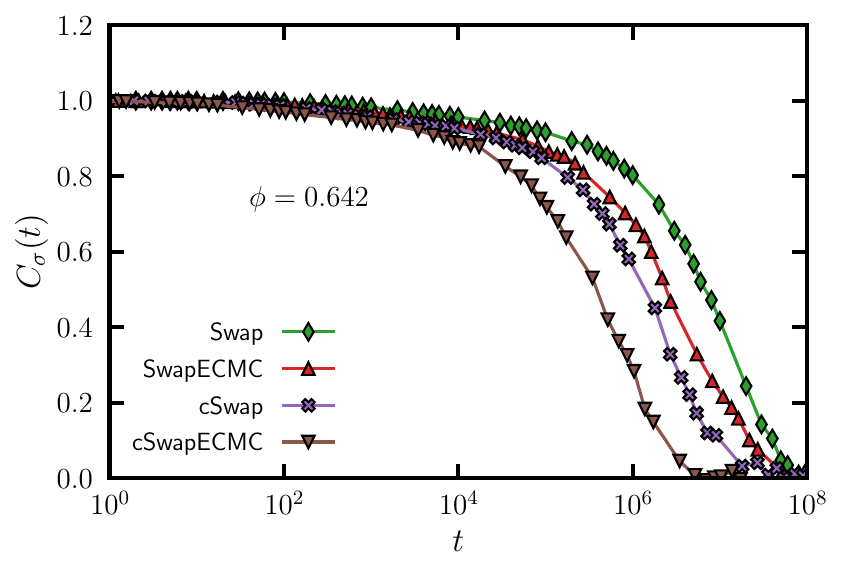}
    \caption{Time relaxation of $C_\sigma(t)$ defined in Eq.~(\ref{eq:diameter}) for Swap, SwapECMC, cSwap and  cSwapECMC algorithms. The dynamics of diameter fluctuations follows the translational dynamics.}
    \label{fig:Csigma} 
\end{figure}

Finally, we track the diameter dynamics by measuring the time autocorrelation function of diameter fluctuations~\cite{ninarello2017models}, defined as  
\begin{equation}
    C_\sigma(t) \equiv \left\langle \frac{\sum_i\delta \sigma_i(t)\delta\sigma_i(0)}{\sum_i \delta\sigma_i(0)^2}\right\rangle ,
    \label{eq:diameter}
\end{equation}
with $\delta\sigma_i(t) \equiv \sigma_i(t) - \overline{\sigma}$. The time evolution of $C_\sigma(t)$ at a fixed packing fraction $\phi=0.642$ is shown in Fig.~\ref{fig:Csigma}. We choose this packing fraction because a full decorrelation can be still be observed for the four Monte Carlo algorithms containing swap moves. 

This figure demonstrates that the hierarchy of algorithmic efficiency studied in Sec.~\ref{sec:tau_alpha} is respected when considering the relaxation dynamics of diameter fluctuations. Interestingly, the decay of $C_\sigma(t)$ is accelerated for SwapECMC with respect to Swap, even though the only difference between the two algorithms lies in the position sector. This confirms the physical idea that the acceleration provided by the swap moves comes from a dynamic interplay of particle translations and diameter fluctuations~\cite{ninarello2017models}, so that accelerating one type of degrees of freedom also speeds up all others. \fg{At the single particle level, however, the coupling between diameter dynamics (such as shown in Fig.~\ref{fig:sigma_t_cswap_vs_Swap_highphi}) and position dynamics is relatively weak, as already demonstrated in Ref.~\cite{ninarello2017models}. We have observed a very similar behavior for cSwap.}

\section{Monte Carlo algorithms to produce jammed sphere packings}

\label{sec:jamming}

All Monte Carlo algorithms can be used to prepare jammed configurations of hard spheres~\cite{berthier2009glass}. Starting from an equilibrated configuration in the fluid at $\phi_\text{init}$, we suddenly apply a very large pressure using $NPT$ simulations combined with MMC, Swap and cSwap algorithms, so that only volume moves with $\delta V < 0$ are accepted. The high applied pressure is $ \beta P\overline{\sigma}^3=10^{12}$, and the maximum step for box moves is $V_\text{max}=5\times 10^{-5} V_\text{init}$, with $V_\text{init}$ the volume of the initial equilibrium configuration at $\phi_\text{init}$. We carry out compressions using the MMC, Swap and cSwap algorithm for a time $t=4\times 10^7$. As the compression proceeds and the packing fraction increases, most of the translational MMC moves and the box moves are rejected, slowing down the compression process. To compensate for this effect, we implement and adaptive step technique: we track the running average ${p_\text{acc}}$ of the acceptance rate for Metropolis translational moves. When this running average drops below a threshold ${p}_\text{th}=0.2$, we rescale the step size for proposed translational volume moves by a factor $2$. The threshold for the running average of the acceptance rate is rescaled as well, making this procedure iterative: $\delta \leftarrow \delta/2$, $\delta V_\text{max} \leftarrow \delta V_\text{max}/2$, ${p}_\text{th} \leftarrow {p}_\text{th}/2$.

\begin{figure}
    \includegraphics[width=\columnwidth]{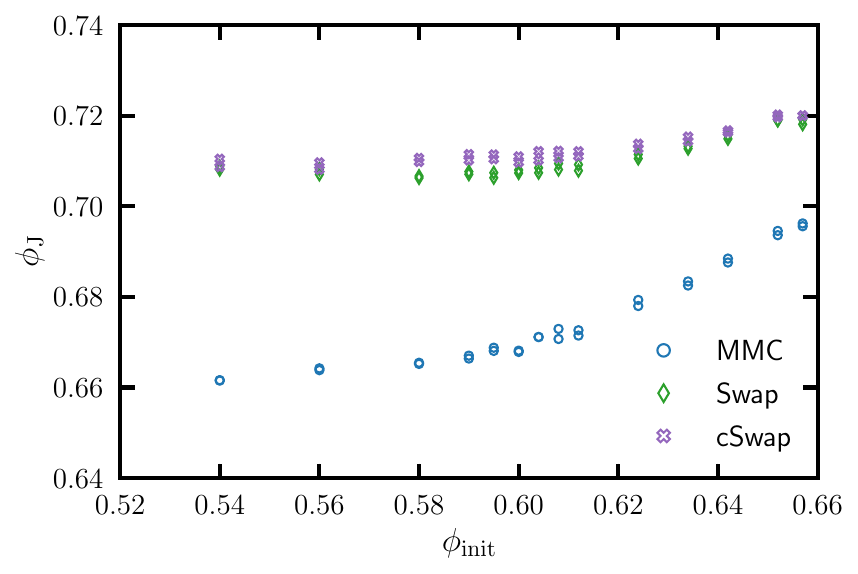}
    \caption{Jamming packing fraction achieved through high-pressure compressions in the NPT ensemble, starting from configurations equilibrated at $\phi_\text{init}$ using MMC, Swap and cSwap.}
    \label{fig:phiJ_phi_init}
\end{figure}

After the pressure quench, the packing fraction of the system slowly increases until saturation to a packing fraction  that corresponds to the jamming point, $\phi_\text{J}$. By construction, $\phi_\text{J}$ may still depend on the initial condition $\phi_{\rm init}$ and on all details of the Monte Carlo algorithm employed to compress the system~\cite{donev2004comment,ohern2004reply,chandler2010dynamics}. 

The evolution of $\phi_\text{J}$ with $\phi_\text{init}$ is shown in Fig.~\ref{fig:phiJ_phi_init}. For conventional MMC, $\phi_\text{J}$ exhibits a plateau at low $\phi_\text{init}$, with a value $\phi_\text{J,pl}^{(\text{MMC})} \approx 0.66$. This value corresponds roughly to the jamming point defined by O'Hern and coworkers~\cite{ohern2002random} for the specific polydisperse model studied here. The jamming packing fraction then increases when $\phi_\text{init}$ grows above $\phi^{*(\text{MMC})}_\text{init} \approx 0.59$. The largest $\phi_\text{J}$ reached with this protocol is $\phi^{(\text{MMC})}_\text{J}\approx 0.6947$, which is consistent with previous results~\cite{ozawa2017exploring}. 

Adding swap moves during the compression gives access to considerably denser jammed packings~\cite{berthier2016equilibrium,brito2018theory,hagh2022transient}. The plateau height of the plateau at low $\phi_{\rm init}$ becomes $\phi_\text{J,pl}^{(\text{Swap})} \approx 0.706$ and is slightly larger for cSwap, 
$\phi_\text{J,pl}^{(\text{cSwap})} \approx 0.709$. It is remarkable that a fast compression with elementary swap moves from a dilute initial condition produces denser packings than a very painful equilibration at large $\phi_{\rm init}$ followed by a compression involving translational moves only~\cite{berthier2016equilibrium}. 

\begin{figure} 
\includegraphics[width=0.9\columnwidth]{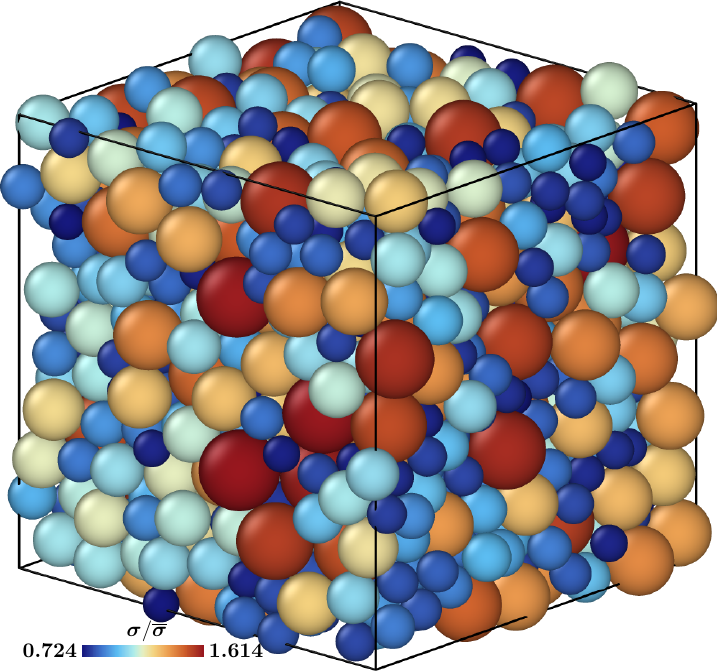}
    \caption{Jammed packing at \fg{$\phi_\text{J} = 0.7202$} obtained combining compressions in the NPT ensemble with the cSwap algorithm.}
    \label{fig:jammed} 
\end{figure}

Combining careful equilibration at larger $\phi_{\rm init}$ to compressions involving swap moves produces even denser jammed configurations, reaching $\phi_{\text{J}}^{\text{Swap}} \approx 0.719$. This value is slightly improved when resorting to the cSwap algorithm during compressions, $\phi_{\text{J}}^{\text{cSwap}} \approx 0.720$. This is significantly larger than in previous efforts for the same model~\cite{ozawa2017exploring}, although this value remains smaller than the close packing density for a periodic crystal of monodisperse spheres, $\phi \simeq 0.7404$. An example of a disordered very dense jammed packing with $\phi_{\rm J} = 0.7202$ is shown in Fig.~\ref{fig:jammed}. 

The small but systematic dependence of  $\phi_\text{J}$ on $\phi_{\rm init}$ in Fig.~\ref{fig:phiJ_phi_init} for Swap and cSwap seems qualitatively different from the results we have obtained in two dimensional hard disks where this dependence was negligible~\cite{ghimenti2024irreversible}. As a result, for two-dimensional systems a rapid compression with swap moves from a dilute or a very glassy initial conditions led to the same jamming packing fraction~\cite{ghimenti2024irreversible,bolton2024ideal}. This is no longer the case in three dimensions, where exploring dense equilibrium states in the fluid helps producing denser jammed packings. 

To confirm this idea, we compare the range of jamming packing fractions obtained using our Monte Carlo methods to the packings obtained by gradient-descent methods combining translational and diameter moves~\cite{brito2018theory,hagh2022transient}. We obtained a jammed polydisperse configuration from the work of Hagh {\it et al.}~\cite{hagh2022transient} which has a different particle size distribution and a polydispersity $\Delta \approx 25 \%$. The augmented gradient-descent protocol yields $\phi_\text{J}^{gd}\approx 0.712$. Using this jammed configuration as a starting point, we first decompress the system to equilibrate the system in the dilute fluid and lose all memory of the initial jammed packing. When we rapidly compress the system from a dilute fluid at $\phi_\text{init}=0.52$ we get 
$\phi_\text{J,pl}^{(\text{MMC})} \approx 0.66$, which corresponds to the jamming point of O'Hern {\it et al}. This becomes $\phi_\text{J,pl}^{(\text{Swap})} \approx 0.710$ when swap moves are introduced in rapid compressions from dilute initial conditions, which is roughly consistent with the gradient-descent method. We then carefully equilibrate the system using $NVT$ simulations and the cSwapECMC algorithm at $\phi_\text{init} = 0.652$ and use these equilibrium configurations as initial conditions for compressions using cSwap. The additional effort spent in preparing a dense equilibrated fluid configuration allows us to reach a jamming packing fraction $\phi_\text{J}^{\text{cSwap}} \approx 0.717$, which is denser than the result obtained from gradient-descent methods alone.  

\section{Conclusion}

\label{sec:conclusion}

The design of efficient sampling algorithms in systems such as glass-formers whose dynamics is controlled by proliferating dynamical bottlenecks has been a challenge on its own since computing hardware has become part of the standard toolbox of theoreticians. When it comes to sampling a set of degrees of freedom according to a given distribution, algorithmic progress is in principle limited by strict mathematical bounds, as discussed in \cite{krauth2021event}, and it is in general simply not known how close one is from the optimally fastest algorithm. This algorithm chimera can be approached on the basis of physical intuition, along with trial and error. But the rules of the game can also be tweaked by introducing out-of-the-box degrees of freedom. This can be done reversibly: this is what the swap algorithm achieves by turning particles into objects of fluctuating sizes. Or this can be done irreversibly, as with event-chain or lifting methods, which exploit the ballistic acceleration of correlated clusters of particles. The present work has presented how a combination of these methods can bring significant speed-up gains even in such uncooperative systems as glass-formers. 

However, much remains to be done, and to begin with, adapting our methods to particles interacting by means of soft continuous potentials stands out as a primary goal. Swap performances do not decrease with continuous potentials~\cite{ninarello2017models}, and some extensions of ECMC to continuous potentials were successful~\cite{michel2014generalized, nishikawa2023liquid}. The most salient challenge will be to design a version of cSwap for glass-formers with continuous potentials to extend the range of applicability of irreversible Monte Carlo methods in the field of supercooled liquids. Yet implementing collective swaps involving finite energy variations will require care so as to preserve global balance. All these perspectives directly follow from our work; they should help the development of efficient and versatile sampling methods for systems with a complex energy landscape. 

\acknowledgments

We thank V. F. Hagh and E. I. Corwin for providing us one of the hard spheres systems studied in this work. We also thank Y. Nishikawa for interesting exchanges on cSwap. We acknowledge the financial support of the ANR THEMA AAPG2020 grant. 

\section*{Author declarations}

\subsection*{Conflicts of interest}

The authors have no conflicts of interest to disclose.

\subsection*{Author contributions}

\textbf{Ludovic Berthier} Conceptualization (equal), Supervision (equal), writing/review \& editing (equal). \textbf{Federico Ghimenti}: Software,  writing/review \& editing (equal). \textbf{Fr\'ed\'eric van Wijland}: Conceptualization (equal), Supervision (equal), writing/review \& editing (equal).

\section*{Data availability}

The data that supports the findings of this study are available from the corresponding author upon reasonable request.

\bibliography{biblio}

\end{document}